\documentclass[12pt]{article}
\usepackage{graphicx}

\newcommand\pubnumber{SNSN-323-63}
\newcommand\pubdate{\today}

\def\Title#1{\begin{center} {\Large #1 } \end{center}}

\newcommand\pubblock{\rightline{\begin{tabular}{l} \pubnumber\\
         \pubdate  \end{tabular}}}
\newenvironment{Abstract}{\begin{quotation}  }{\end{quotation}}
\newenvironment{Presented}{\begin{quotation} \begin{center} 
             PRESENTED AT\end{center}\bigskip 
      \begin{center}\begin{large}}{\end{large}\end{center} \end{quotation}}
\def\Acknowledgements{\bigskip  \bigskip \begin{center} \begin{large}
             \bf ACKNOWLEDGEMENTS \end{large}\end{center}}
\def\beq{\begin{equation}}
\def\eeq#1{\label{#1}\end{equation}}
\def\eeqn{\end{equation}}

\def\beqa{\begin{eqnarray}}
\def\eeqa#1{\label{#1}\end{eqnarray}}
\def\eeqan{\end{eqnarray}}

\let\bar=\overbar

\def\Dslash{\not{\hbox{\kern-4pt $D$}}}
\def\dslash{\not{\hbox{\kern-2pt $\del$}}}

\def\msb{{\bar{\ssstyle M \kern -1pt S}}}

\begin{document}
\hyphenation{sy-ste-ma-tic}
\hyphenation{me-a-su-re-ment}
\begin{titlepage}
\pubblock

%\vfill
\Title{First Results from the TOTEM Experiment}
%\vfill
%
%\Author{Giuseppe Latino \support}
%\Author{Giuseppe Latino}
%\Address{\Siena}
%
\hyphenation{Ni-e-wi-a-dom-ski}

G.~Latino$^{7b,}$\footnote{Speaker. Corresponding author. Phone: +39-050-2214439. E-mail: 
giuseppe.latino@pi.infn.it.},
G.~Antchev$^{*}$, P.~Aspell$^{8}$, I.~Atanassov$^{8,}$$^{*}$, V.~Avati$^{8}$, J.~Baechler$^{8}$,
V.~Berardi$^{5b,5a}$, M.~Berretti$^{7b}$, E.~Bossini$^{7b}$, M.~Bozzo$^{6b,6a}$, P.~Brogi$^{7b}$ ,
E.~Br\"{u}cken$^{3a,3b}$, A.~Buzzo$^{6a}$, F.~Cafagna$^{5a}$, M.~Calicchio$^{5b,5a}$,
M.~G.~Catanesi$^{5a}$, C.~Covault$^{9}$, T.~Cs\"{o}rg\H{o}$^{4}$, M.~Deile$^{8}$,
K.~Eggert$^{9}$, V.Eremin$^{*}$, R. Ferretti$^{6a, 6b}$, F.~Ferro$^{6a}$,
A.~Fiergolski$^{*}$, F.~Garcia$^{3a}$, S.~Giani$^{8}$,
V.~Greco$^{7b,8}$, L.~Grzanka$^{8,}$$^{*}$, J.~Heino$^{3a}$, T.~Hilden$^{3a,3b}$, M.R.~Intonti$^{5a}$,
J.~Ka\v{s}par$^{1a,8}$, J.~Kopal$^{1a,8}$, V.~Kundr\'{a}t$^{1a}$, K.~Kurvinen$^{3a}$,
S.~Lami$^{7a}$, R.~Lauhakangas$^{3a}$, T. Leszko$^{*}$, E.~Lippmaa$^{2}$,
M.~Lokaj\'{\i}\v{c}ek$^{1a}$, M.~Lo~Vetere$^{6b,6a}$, F.~Lucas~Rodr\'{\i}guez$^{8}$,
M.~Macr\'{\i}$^{6a}$, L.~Magaletti$^{5b,5a}$, A.~Mercadante$^{5b,5a}$, S.~Minutoli$^{6a}$, 
F.~Nemes$^{4,}$$^{*}$, 
H.~Niew- iadomski$^{8}$,
E.~Oliveri$^{7b}$, F.~Oljemark$^{3a,3b}$, R.~Orava$^{3a,3b}$, M.~Oriunno$^{8}$$^{*}$,
K.~\"{O}sterberg$^{3a,3b}$, P.~Palazzi$^{7b}$,
J.~Proch\'{a}zka$^{1a}$, M.~Quinto$^{5a}$, E.~Radermacher$^{8}$, E.~Radicioni$^{5a}$,
F.~Ravotti$^{8}$, E.~Robutti$^{6a}$, L.~Ropelewski$^{8}$, G.~Ruggiero$^{8}$,
H.~Saarikko$^{3a,3b}$, G.~Sanguinetti$^{7a}$,  A.~San- troni$^{6b,6a}$,
A.~Scribano$^{7b}$, W.~Snoeys$^{8}$, J.~Sziklai$^{4}$, C.~Taylor$^{9}$,
N.~Turini$^{7b}$, V.~Vacek$^{1b}$, M.~V\'{i}tek$^{1b}$, J.~Welti$^{3a,b}$, J.~Whitmore$^{10}$.
\begin{center}
\small{(TOTEM Collaboration)}
\end{center}
\small{
$^{1a}${Institute of Physics, Academy of Sciences of the Czech Republic, Praha, Czech Republic.}
$^{1b}${Czech Technical University, Praha, Czech Republic.}
$^{2}${National Institute of Chemical Physics and Biophysics NICPB, Tallinn, Estonia.}
$^{3a}${Helsinki Institute of Physics, Finland.}
$^{3b}${Department of Physics, University of Helsinki, Finland.}
$^{4}${MTA KFKI RMKI, Budapest, Hungary.}
$^{5a}${INFN Sezione di Bari, Italy.}
$^{5b}${Dipartimento Interateneo di Fisica di Bari, Italy.}
$^{6a}${Sezione INFN, Genova, Italy.}
$^{6b}${Universit\`{a} degli Studi di Genova, Italy.}
$^{7a}${INFN Sezione di Pisa, Italy.}
$^{7b}${Universit\`{a} degli Studi di Siena and Gruppo Collegato INFN di Siena, Italy.}
$^{8}${CERN, Geneva, Switzerland.}
$^{9}${Case Western Reserve University, Dept. of Physics, Cleveland, OH, USA.}
$^{10}${Penn State University, Dept. of Physics, University Park, PA, USA.}
$^{*}${Visitor from an external Institution.}
}
~~\\
%\vfill
%\vskip 0.4cm
%
\begin{Abstract}
The first physics results from the TOTEM experiment are here reported, concerning the 
measurements of the total, differential elastic, elastic and inelastic pp cross-section 
at the LHC energy of $\sqrt{s}$ = 7 TeV, obtained using the luminosity measurement from CMS. 
A preliminary measurement of the forward charged particle $\eta$ distribution is also shown.
\end{Abstract}
\vfill
\begin{Presented}
MPI@LHC 2010\\
Glasgow, UK, November 29 -- December 3, 2010.
\end{Presented}
%\vfill
\end{titlepage}

\section{Introduction}

The TOTEM experiment~\cite{TOTEM_TDR}, sharing with CMS~\cite{CMS} the same interaction point IP5 
of the CERN LHC, is one of the six experiments that investigate high energy 
physics at this new machine. It has been designed for the measurement of the total $pp$ cross-section
($\sigma_{tot}$) with a precision down to 1$\div$2\,$\%$, for the study of the 
nuclear elastic $pp$ differential cross-section ($d{\sigma}_{el} / dt$) over a wide range 
of the squared four-momentum transfer $|t| \sim (p\theta)^2$ ($\sim 10^{-3} < |t| < 
10\,{\rm GeV}^{2}$) and for the development of a 
comprehensive physics programme on diffractive dissociation, partially in cooperation 
with CMS. These studies will allow to distinguish among different models of 
soft proton interactions, giving a deeper understanding of the proton structure. 

Due to the large uncertainties on available high energy data, the theoretical 
predictions for $\sigma_{tot}$, based on fits to existing measurements
on $pp$ and $p\bar p$ scattering according to different models, are 
typically in the 90$\div$130 mb range~\cite{COMPETE}. The TOTEM measurement of 
$\sigma_{tot}$ at the level of 1$\div$2\,$\%$ will allow to discriminate among different 
models. In order to reach such a small error, the measurement will be based on the 
``luminosity independent method'' which, combining the optical theorem with the total rate, 
gives $\sigma_{tot}$ (but also the machine luminosity $\mathcal{L}$, useful for calibration 
purposes) in terms of the total elastic rate, the total inelastic one and the differential 
elastic cross-section extrapolated to $t = 0$ (optical point)~\cite{TOTEM_JINST}. 
The uncertainty on this extrapolation depends on the acceptance for protons scattered 
at small $|t|$ values, hence at small angles. This requires a small beam angular 
divergence at the IP, which can be achieved in special runs with high $\beta^*$ 
machine optics and typically low $\mathcal{L}$ (in order to have a negligible pile up). 
An approved optics with $\beta^* =$ 1540 m is expected to give a $\sigma_{tot}$ 
($\mathcal{L}$) measurement at the level of 1$\div$2\,$\%$ (2\%), with the 
systematic uncertainty dominated by the uncertainty on the corrections to trigger losses 
for low mass Single Diffraction events~\cite{TOTEM_JINST}.
The $\beta^* =$ 90 m optics, achievable without modifying the standard LHC 
injection optics and already tested, can allow a preliminary
$\sigma_{tot}$ measurement with a higher systematic uncertainty 
dominated by the extrapolation to $t$ = 0~\cite{TOTEM_JINST}.

Many details of diffractive (due to colour singlet exchange) and non-diffractive (due to 
colour exchange) inelastic interactions are still poorly understood. At the same time, these
processes, with close ties to proton structure and low-energy QCD, represent a big fraction 
of $\sigma_{tot}$. The majority of diffractive events exhibits intact (``leading'') protons
characterized by their $t$ and fractional momentum loss $\xi \equiv \Delta 
p/p$. TOTEM is able to measure $\xi$-, $t$- and mass-distributions with 
acceptances depending on the beam optics. Furthermore, the charged particle flow 
in the forward region is studied in TOTEM, with the aim to provide in particular 
a significant contribution to the understanding of cosmic ray physics. The existing models 
give in fact predictions on energy flow, multiplicity and other quantities related to cosmic 
ray air showers, with significant inconsistencies in the forward region. 

The integration of TOTEM with the CMS detector is also foreseen, resulting in the 
largest acceptance detector ever built at a hadron collider. This will offer the 
possibility of more detailed studies on inelastic events, 
including hard diffraction~\cite{CMS_TOTEM_TDR}.

In order to fulfill its physics programme TOTEM has to cope with the challenge of triggering
and recording events in the very forward region with a good acceptance for particles produced
at very small angles with respect to the beam. This involves the detection of
elastically scattered and diffractive protons at a location very close to the beam, together with
efficient forward charged particle detection in inelastic events with losses reduced to few per-cents.

\section{Detector Components}

\begin{figure}[htb]
%\centering
\includegraphics[height=1.8in, width=10.5cm]{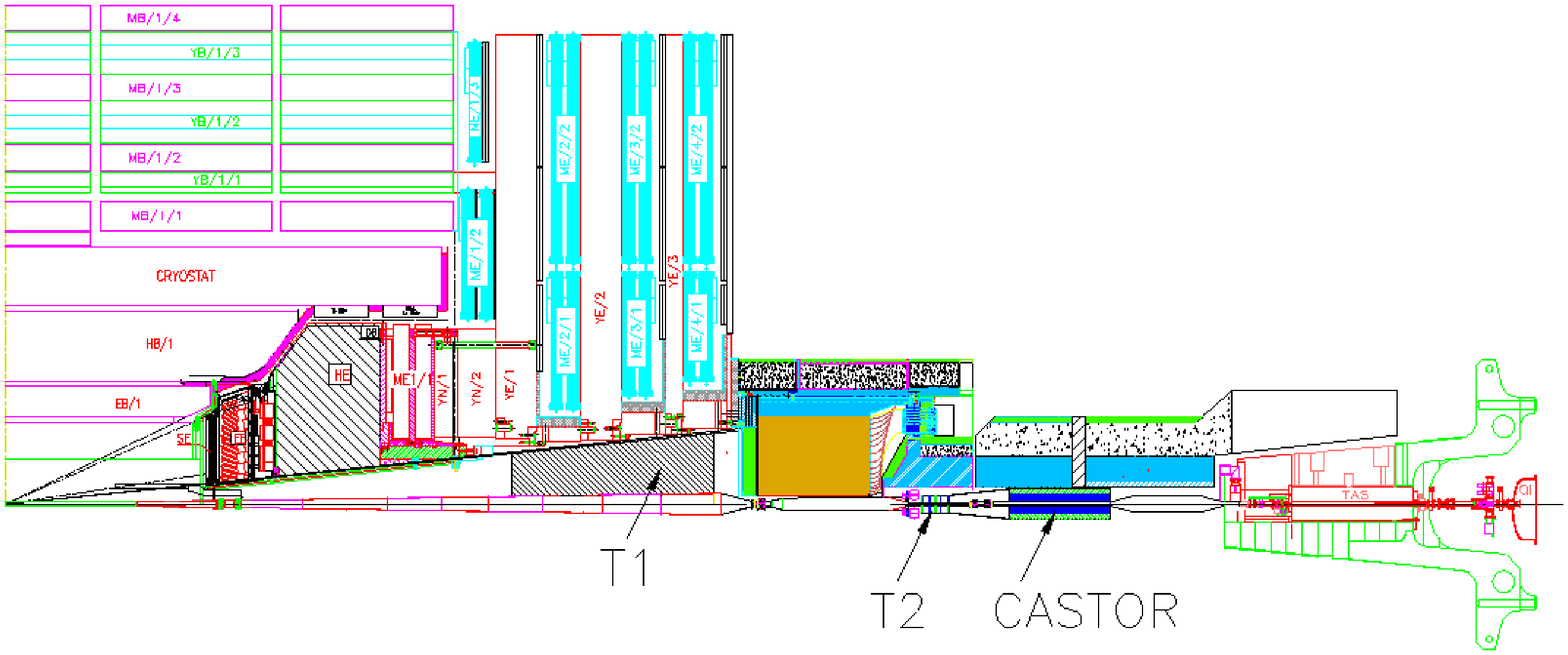}
\vskip -4.5cm
\hskip 5.0cm
\includegraphics[height=0.8in, width=10.5cm]{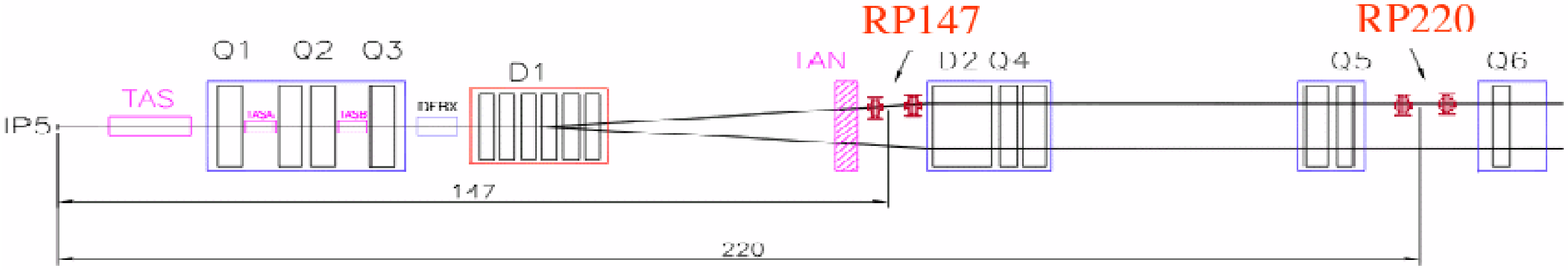}
\vspace{1.5cm}
\caption{Top: Roman Pots location along the LHC beam-line.
Bottom: T1 and T2 location in the forward region of the CMS detector.
All TOTEM detectors are located on both sides of IP5.}
\label{fig:totem_exp}
\end{figure}

The TOTEM experiment includes three detector components located on both sides of the
interaction point IP5 (Figure~\ref{fig:totem_exp}): the T1 and T2 inelastic telescopes, 
embedded inside the forward region of CMS, and the ``Roman Pots'' (RPs) detectors, 
placed on the beam-pipe of the outgoing beam in two stations at about 147 m and 220 m from IP5. 

\begin{figure}[htb]
\centering
\includegraphics[height=1.5in, width=0.32\linewidth]{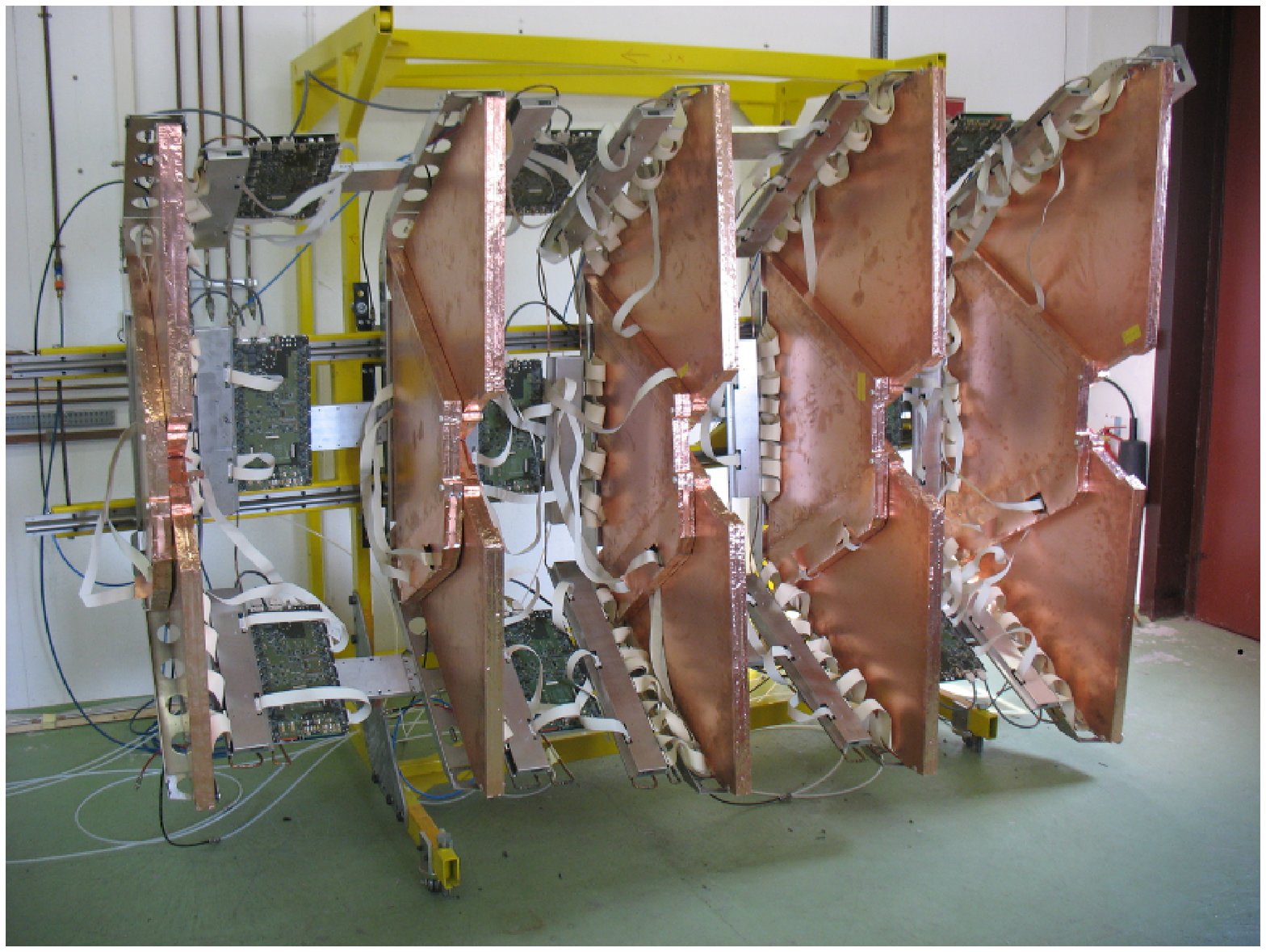}
\includegraphics[height=1.5in, width=0.32\linewidth]{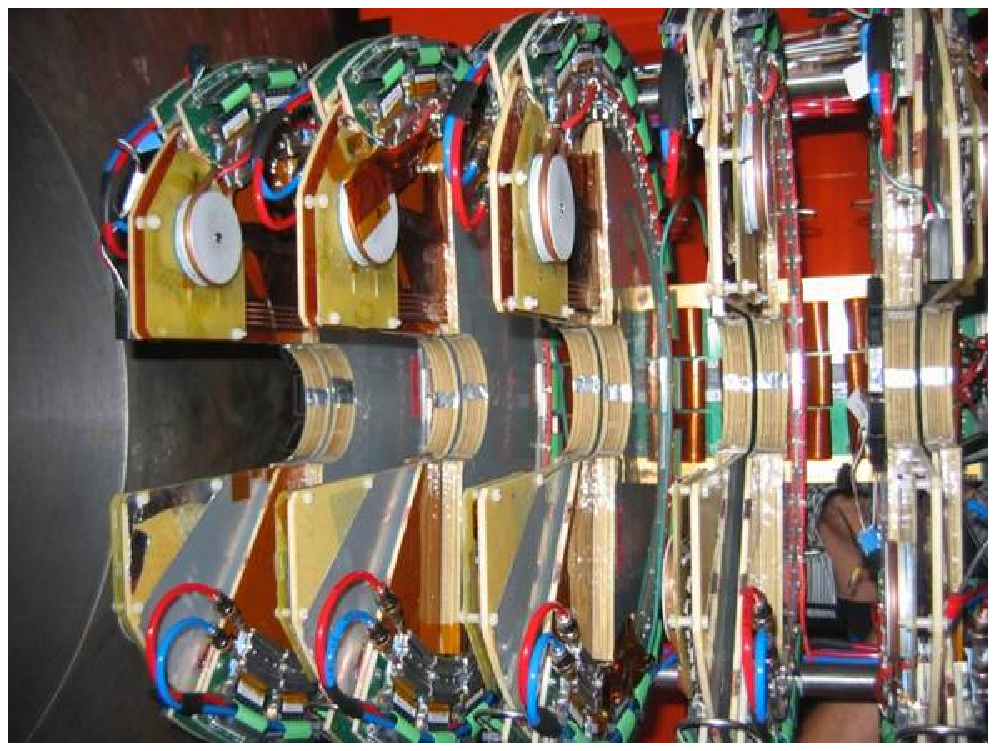}
\includegraphics[height=1.5in, width=0.32\linewidth]{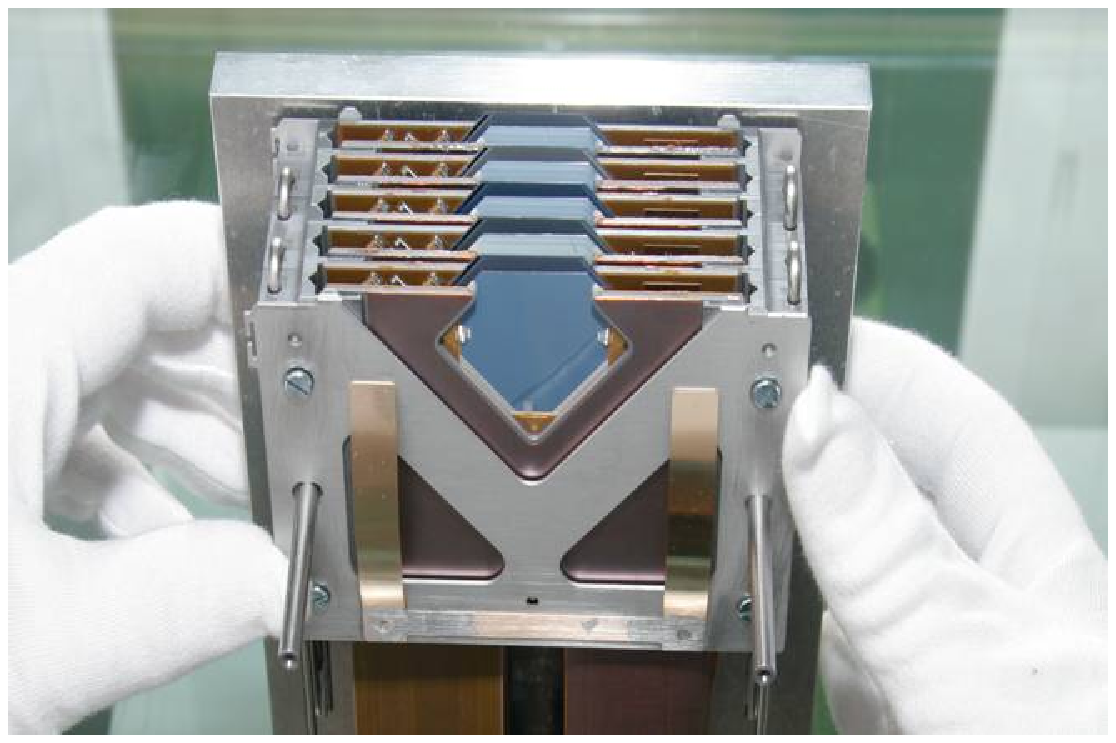}
\caption{Left (Centre): one half-arm of the T1 (T2) telescope. 
Right: silicon detectors hosted in one pot.}
\label{fig:totem_det}
\end{figure}

T1 and T2 are gas detectors providing charged track reconstruction
in the 3.1 $<$ $|\eta|$ $<$ 6.5 range ($\eta = -ln(tan\frac{\theta}{2})$), with a
2$\pi$ coverage and with a very good efficiency~\cite{CMS_TOTEM_TDR}.
Their trigger capability, with an acceptance grater than 95$\%$ for all inelastic events,
allows the measurement of inelastic rates with small losses. At the same time they are
used for the reconstruction of the event interaction vertex, allowing to reject
background events.
Located at $\sim$ 9 m from IP5, each T1 telescope arm covers the range 3.1 $<$ $|\eta|$ $<$ 4.7
and consists of five planes formed by six trapezoidal ``Cathode Strip Chambers''
(CSC)~\cite{TOTEM_TDR} (Figure~\ref{fig:totem_det}, left).
These CSCs, with 10 mm thick gas gap and a gas mixture of Ar/CO$_2$/CF$_4$ ($40\%/50\%/10\%$),
give three measurements of the charged particle coordinates with a spatial resolution
of $\sim$ 1 mm: anode wires (pitch of 3 mm), also giving level-1 trigger information, are parallel
to the trapezoid base; cathode strips (pitch of 5 mm) are rotated by $\pm$ $60^o$
with respect to the wires.
The T2 telescope, based on ``Gas Electron Multiplier'' (GEM) technology~\cite{GEM},
extends charged track reconstruction to the range 5.3 $<$ $|\eta|$ $<$
6.5~\cite{TOTEM_TDR}. Each half-arm, located at $\sim$ 13.5 m from IP5, is
made by the combination of ten aligned detectors planes having an almost semicircular
shape (Figure~\ref{fig:totem_det}, center). This novel gas detector technology is optimal
for the T2 telescope thanks to its good spatial resolution, excellent rate capability and good
resistance to radiation. The T2 GEMs are characterized by a triple-GEM structure and a
gas mixture of Ar/CO$_2$ (70$\%$/30$\%$)~\cite{TOTEM_TDR}.
The read-out board has two separate layers with different patterns: one with 256x2 concentric
circular strips (80\,$\mu$m wide, pitch of 400\,$\mu$m), allowing track radial coordinate
reconstruction with a resolution of $\sim$ 100\,$\mu$m; the other with a
matrix of 24x65 pads (from 2x2\,mm$^2$ to 7x7\,mm$^2$ in size) providing level-1 trigger
information and track azimuthal coordinate reconstruction.

The RPs are special movable beam-pipe insertions designed to detect ``leading''
protons with a scattering angle down to few $\mu$rad. They host silicon detectors
which are moved very close to the beam when it is in stable conditions.
Each RP station is composed of two units in order to have a lever harm for a better local track
reconstruction and a higher efficiency of trigger selection by track angle. Each unit 
consists of three pots,
two vertical and one horizontal completing the acceptance for diffractively scattered protons.
Each pot contains a stack of 10 planes of silicon strip detectors (Figure~\ref{fig:totem_det}, left).
Each plane has 512 strips (pitch of 66 $\mu$m), oriented at $+45^o$ (5 planes) or at $-45^o$ 
(5 planes) w.r.t. the detector edge facing the beam, allowing a single hit resolution of 
$\sim$ 20 $\mu$m.
As the detection of protons elastically scattered at angles down to few $\mu$rads requires a
detector active area as close to the beam as $\sim$ 1 mm, a novel ``edgeless planar silicon''
detector technology has been developed for TOTEM RPs in order to have an edge dead zone minimized to
only $\sim$ 50 $\mu$m~\cite{RP_Silicon}.

The read-out of all TOTEM sub-detectors is based on the digital VFAT chip, specifically designed for 
TOTEM and characterized by trigger capabilities~\cite{TOTEM_JINST}.
\section{First Physics Results}

\begin{figure}[htb]
\centering
\hskip -0.4cm
\includegraphics[height=2.36in,width=0.535\linewidth]{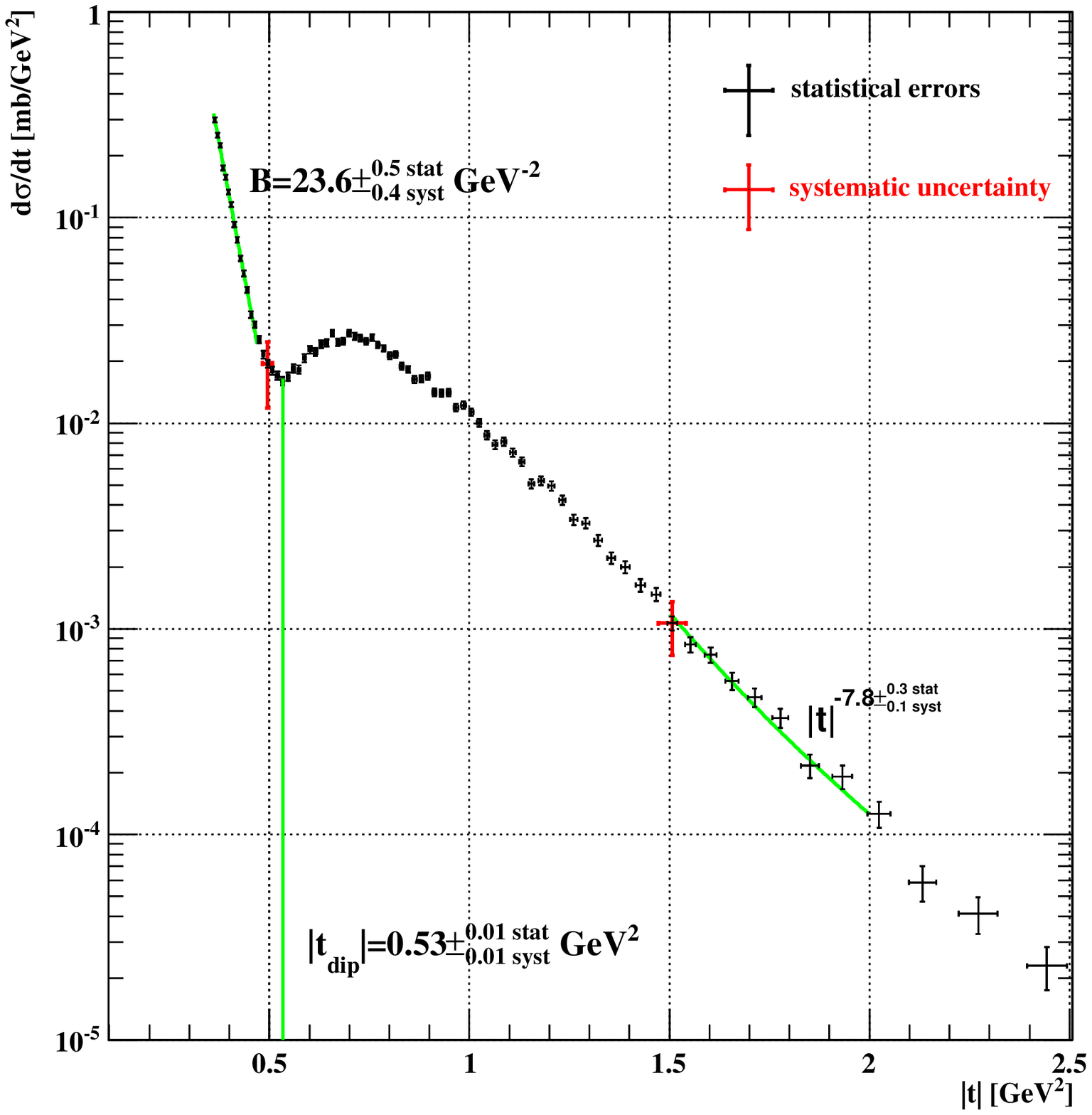}
\hskip -0.5cm
\includegraphics[height=2.2in,width=0.515\linewidth]{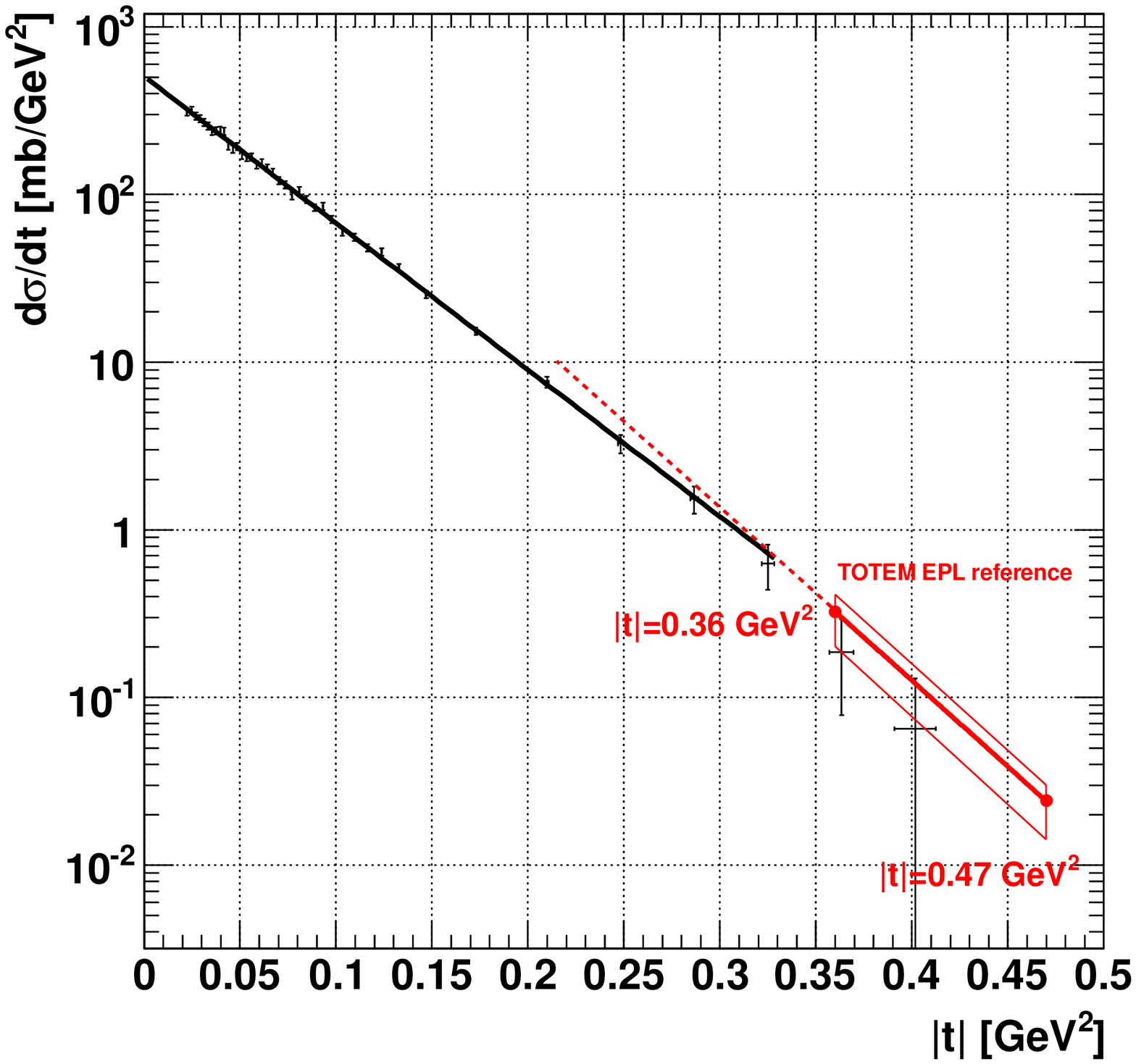}
\caption{Left: first measurement of $d{\sigma}_{el}/dt$ with the standard 
$\beta^* =$ 3.5 m LHC optics. Right: extension of the measurement at low $|t|$ 
with the $\beta^* =$ 90 m optics.
}
\label{fig:dSigmaEl_dt}
\end{figure}

TOTEM has performed a first measurement of $d{\sigma}_{el} / dt$ at $\sqrt{s}$ = 7 TeV in the 
0.36 $<$ $|t|$ $<$ 2.5 GeV$^2$ range using data taken in 2010 with the standard optics 
($\beta^* =$ 3.5 m) during a dedicated run at low luminosity~\cite{EPL_SigEl}. A total 
luminosity of 6.1 nb$^{-1}$ was integrated with the RP detectors approaching the beams as close 
as 7 times the transverse beam size (${\sigma}_{b}$). The background was significantly 
reduced a the trigger level by requiring collinear hits in a pot in at least three of the 
five planes for each projection (trigger tracks) on both sides of the IP. 
Elastic candidates were then selected offline by applying proper cuts in order to reject the 
background from diffractive events and by requiring a reconstructed track in both 
projections of the vertical RP units on each side of the IP in a ``diagonal'' topology: top (bottom) 
left of IP - bottom (top) right of IP.
Dedicated procedures have been performed in order to ensure the precision and the reproducibility of 
all RP detector planes alignment with respect to each other and to the position of the beam centre, 
one of the most delicate ad difficult tasks of the experiment. 
A precise relative alignment (at the 10 $\mu$m level) of all three RPs in a unit has been obtained during 
the measurement by correlating their position via common particle tracks reconstruction in the overlap zone 
of the horizontal RPs with the vertical ones. The global symmetrical alignment of all the RPs with respect 
to the beam centre has been obtained (with a precision $\sim$ 50 $\mu$m) during a dedicated beam fill by moving 
them towards the sharp beam edge cut by the beam collimators, until a beam losses spike was observed 
downstream of the RPs. The final horizontal and vertical alignment has then been 
achieved from studies on the reconstructed tracks. 
The horizontal (${\Theta}_x^*$) and vertical (${\Theta}_y^*$) scattering angles at the IP were deduced 
from the measurement of the track angle (for ${\Theta}_x^*$) and of the displacement in $y$ (for 
${\Theta}_y^*$) at the RP stations using the optical functions, which describe the explicit proton 
path through the LHC magnetic elements as a function of the proton position and scattering angle at the IP. 
Event selection has been performed by requiring a strict correlation (consistent with the beam divergence 
at the IP) 
between the two ${\Theta}_x^*$ (and ${\Theta}_y^*$) reconstructed on both sides of IP, resulting in a 
t-resolution of $\delta t$ = 0.1 GeV$\sqrt{|t|}$.
The detector efficiency and acceptance corrections have been computed from simulation, while bin migration 
due to resolution and beam divergence effects has been corrected using two 
independent unfolding procedures based on analytical and on MC methods which gave 
consistent results. The contribution of background events passing the selection cuts was 
evaluated from studies on data. 
The total luminosity associated to the collected data has been derived from the instantaneous 
luminosity measurement performed by CMS with an uncertainty of 4$\%$~\cite{CMS_lumi1,CMS_lumi2}, 
integrated over the 
data taking period and then corrected for trigger 
and 
DAQ efficiency effects.
Figure~\ref{fig:dSigmaEl_dt} (left) shows the measured $d{\sigma}_{el} / dt$ 
with the related statistical errors, the one on $t$ given by the beam divergence. 
The systematic uncertainties, reported as reference only in two points, 
are dominated in $t$ by optics and alignment, while in $d{\sigma}_{el} / dt$ by the uncertainty on the 
efficiency corrections and the resolution unfolding. 
For $|t|$ $<$ 0.47 GeV$^{2}$ the data can be described by an exponential function with 
slope $B = (23.6 \pm 0.5^{stat} \pm 0.4^{syst})$ GeV$^{-2}$, which is expected to change 
at smaller $|t|$ values. The expected diffractive minimum, typically pronounced in pp scattering, is then observed 
at $|t| = (0.53 \pm 0.01^{stat} \pm 0.01^{syst})$ GeV$^{2}$. At higher $|t|$ the $d{\sigma}_{el} / dt$ 
can be described by a power law $|t|^{-n}$, with $n = 7.8 \pm 0.3^{stat} \pm 0.1^{syst}$, in the 
$|t|$-range 1.5 - 2.0 GeV$^2$. The comparison with the predictions of different available theoretical 
models shows a partial consistency with the data (slope $B$, dip position and exponent $n$ at large $|t|$) 
only for some of them.    

Analyzing the data from a short 2011 run with dedicated large $\beta^*$ optics ($\beta^* =$ 90 m) 
and low luminosity, TOTEM has also measured $d{\sigma}_{el}/dt$ at low $|t|$ in the 
$2\times 10^{-2} < |t| < 0.42$ GeV$^2$ range~\cite{EPL_SigEl_SigTOT}. 
During this special run a total luminosity of 1.65 $\mu$b$^{-1}$ was integrated with 
the RP detectors placed 10 times ${\sigma}_{b}$ from the beam centre, using a loose trigger requiring 
a track segment in any of the vertical RPs in at least one of the two transverse 
projections. The analysis strategies were substantially the same as in the 
previous measurement at larger $|t|$, the luminosity being still provided by the CMS 
experiment with an uncertainty of 4$\%$. The uncertainty on the $|t|$ scale for this optics 
was found to vary from 0.8$\%$ at low $|t|$ to 2.6$\%$ at high $|t|$. 
The results are reported in Figure~\ref{fig:dSigmaEl_dt} (right), were the exponential 
fit at the lower end of the $|t|$ range of the previous measurement is 
also shown for comparison, showing a very good agreement between the two measurements 
performed with different optics.
An exponential fit with a slope $B = (20.1 \pm 0.2^{stat} \pm 0.3^{syst})$ GeV$^{-2}$
describes the lowest range of $|t|$ from 0.02 to 0.33 GeV$^2$. The low $|t|$ value reached 
with this optics made the exponential extrapolation to $t$ = 0 possible, allowing 
the first measurement of the total pp cross-section at the LHC using the 
optical theorem, the luminosity measurement from CMS and the $\rho$ parameter from theoretical 
predictions~\cite{COMPETE}. 
A $\sigma_{tot}$ of $(98.3 \pm 0.2^{stat} \pm 2.8 ^{syst})$ mb was
obtained, which is in good agreement with the expectation from the overall fit of previously
measured data over a large range of energies~\cite{COMPETE}. The errors on this measurement 
are dominated by the extrapolation to $t$ = 0 and by the luminosity uncertainty.
Furthermore, the integration of $d{\sigma}_{el} / dt$ gave a 
$\sigma_{el}$ of $(24.8 \pm 0.2^{stat} \pm 1.2^{syst})$ mb. 
By combining the $\sigma_{tot}$ and $\sigma_{el}$ measurements, a inelastic 
cross-section ($\sigma_{inel}$) of  
$(73.5 \pm 0.6^{stat} ~^{+1.8}_{-1.3} ~^{syst})$ mb was inferred, which is in good agreement 
with the measurements of the ALICE~\cite{SigInel_ALICE}, ATLAS~\cite{SigInel_ATLAS} and 
CMS~\cite{SigInel_CMS} experiments within the quoted errors. 
\begin{figure}[htb]
\centering
\includegraphics[height=2.7in]{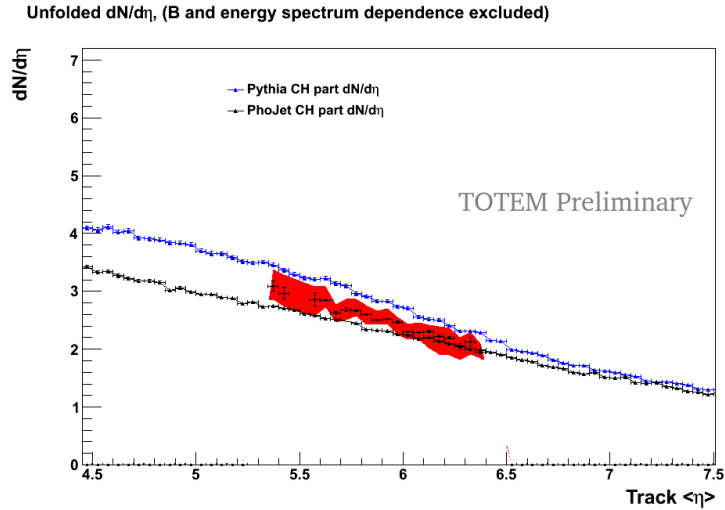}
\caption{Preliminary measurement of $\frac{dN}{d{\eta}}$ (black points with statistical error), 
compared to predictions from Pythia6 (blue triagles) and Phojet (black triangles). The red band 
represents the combination of the main systematic uncertainties.}
\label{fig:dN_dEta}
\end{figure}

A preliminary measurement of the forward charged particle $\eta$ distribution has 
also been performed by TOTEM using the data taken in special 2011 runs at low luminosity 
with an inclusive T2 trigger. Particular effort has been devoted to correct for 
misalignment biases, found to be dominated by global T2 quarter displacements.
The relative alignment between the two quarters of an arm has been obtained using tracks 
reconstructed in the overlap regions, while the global alignment has been derived 
by studying the expected symmetry in the track parameters distributions
and the position on each T2 plane of the ``beam pipe shadow'' (very low track 
efficiency radial zone due to primary particles absorbed by the $\eta \sim $ 5.54 beam pipe cone). 
Secondary track rejection has been derived from data analysis, 
while primary track efficiency and smearing effects correction have been obtained 
from MC studies.
The results are reported in Figure~\ref{fig:dN_dEta}, where
the black points show the experimental measurements with their statistical error. 
The red band represents the sum in quadrature of the main systematic uncertainties, 
related to the estimation of the track efficiency and detector alignment 
corrections and to the subtraction of secondary track contribution.
Work is in progress for the determination of the residual systematics related to the MC 
modeling of the forward particle energy spectrum and to the simulation of the magnetic 
field effects. These last, less relevant 
uncertainties, are expected to give an additional contribution at the level of few percent.   
\section{Summary and Conclusions}

The TOTEM detectors have been completely installed and, after a commissioning period, 
are fully operative. From the analysis of the data taken during dedicated runs at low 
luminosity with both the standard $\beta^* =$ 3.5 m and the high $\beta^* =$ 90 m optics, 
the first (luminosity-dependent) measurements of the total, differential elastic and elastic 
pp cross-section at the LHC energy of $\sqrt{s}$ = 7 TeV have been obtained. The inelastic pp 
cross-section has also been inferred from the total and elastic ones, which is in 
good agreement within the errors with the measurements of other LHC experiments 
(ALICE, ATLAS and CMS).  
A preliminary measurement of the forward charged particles $\eta$ distribution has 
also been performed.

The inclusion of the T1 and T2 inelastic telescopes in the analysis of the data 
to be taken during a high statistics run with $\beta^* =$ 90 m, foreseen in fall 2011,       
will allow a luminosity-independent measurement of the total pp cross-section and a 
detailed study of low mass diffraction.
\Acknowledgements

We are grateful to the Conference Organizers for their kind invitation to 
this nice appointment in the magic atmosphere of Glasgow.
\end{document}